\RequirePackage{lineno}
\documentclass[preprint,showpacs,preprintnumbers]{revtex4}

\newcommand{\wprime}{\ensuremath{W^\prime}~}
\newcommand{\wprimep}{\ensuremath{W^{\prime+}}}
\newcommand{\wprimem}{\ensuremath{W^{\prime-}}}
\newcommand{\Tau}{\ensuremath{\tau_h}}
\newcommand{\tauTau}{\ensuremath{\tau_h\tau_h}}
\newcommand{\lepTau}{\ensuremath{\ell\tau_h}}
\newcommand{\pt}{\ensuremath{p_T}}
\newcommand{\mttwo}{\ensuremath{M_{T2}}}
\newcommand{\MET}{\ensuremath{p_T^{miss}}}
\newcommand{\SumMT}{ \ensuremath{\Sigma M_\mathrm{T}^{\tau_i}}~}
\newcommand{\gR}{\ensuremath{g^\prime_R}}
\newcommand{\gL}{\ensuremath{g^\prime_L}}
\newcommand{\gSM}{\ensuremath{g_{\scriptscriptstyle SM}}}

\usepackage{graphicx}
\usepackage{dcolumn}
\usepackage{bm}
\usepackage{epsfig}
\usepackage{epstopdf}
\usepackage{grffile}
\usepackage{color}
\usepackage{colordvi}
\usepackage{amssymb}
\usepackage{rotating}
\usepackage{lscape}
\usepackage{float}

\usepackage{footnote}
 \makesavenoteenv{environmentname}

\usepackage{hyperref}

\usepackage[T1]{fontenc} 

\begin{document}

\title{$W'$ Pair Production in the Light of CMS Searches}

\author{Saeid Paktinat Mehdiabadi}
\email{paktinat@ipm.ir}
\affiliation{School of Particles and Accelerators, Institute for Research in Fundamental Sciences (IPM), P.O.Box 19395-5531, Tehran, Iran\\
Faculty of Physics, Yazd University, P.O. Box 89195-741, Yazd, Iran}

\author{Leila Zamiri}
\email{leila.zamiri@gmail.com}
\affiliation{Independent Researcher}

\date{\today}

\begin{abstract}
For the first time, the pair production of the heavy charged gauge bosons, known as $W'$ bosons is considered, when both decay to $\tau$ leptons. The reported detailed efficiency of object/event selection by the CMS experiment is used to find the lower limit on the mass of $W'$ boson. Various assumptions for the coupling of the new gauge boson are examined and the results are reported. In the case of a SM-like $W'$ boson, masses below 290 GeV are excluded at 95\% confidence level. The method can be used to constrain other new models with  similar final state.
\end{abstract}

\maketitle

\section{Introduction}\label{sec:int} 
New heavy charged gauge bosons, called \wprime bosons, are predicted by numerous different extensions of the standard model of the elementary particles (SM). 
The SM $W$ boson couples only to left-handed fermions, whereas the coupling of \wprime boson can be completely left-handed, right-handed or a mixture of both. 

The general form of the lagrangian describing the fermionic interactions of \wprime boson is given in  Ref.~\cite{Sullivan:2002jt}
\begin{eqnarray}
{\cal L}& =& \frac{V_{ij}}{2\sqrt{2}}\bar{f_i} \gamma_{\mu}(g^\prime_{R} (1+{\gamma}^5)+
g^\prime_{L}
(1-{\gamma}^5)) W^{\prime \mu} f_j  \nonumber\\
&+& \mathrm{h.c.},
\label{eq:lagrangian}
\end{eqnarray}
where $g'_{R(L)}$ are the right handed (left-handed) coupling constants. The $V_{ij}$ matrix refers to a $3\times3$ identity matrix for leptons and the CKM matrix for quarks. The $(1\pm{\gamma^5})$ operators represent left and right-handed chiral projection operators. In the case, \gR = 0 and \gL $\neq$ 0 (pure left-handed), both leptons and quarks can couple to \wprime boson, but where \gR $\neq$ 0 and \gL = 0 (pure right-handed) only quarks can couple to \wprime boson, because we either do not introduce right-handed neutrinos or they are assumed to be much heavier than \wprime boson. 

In this paper, for the first time, we consider a situation where two opposite-sign \wprime bosons are produced. Since nowadays the colliders center of mass energy is sufficiently high, such processes can be accessible. Due to the important role of the third generation fermions in many new physics scenarios, each \wprime boson is decayed to a $\tau$ lepton and its neutrino ($\nu_{\tau}$). Since we ask for two $\tau$ leptons in the final state, $g'_L$ can not be zero. 

\begin{figure}[!htb]
	\includegraphics*[width=.45\textwidth]{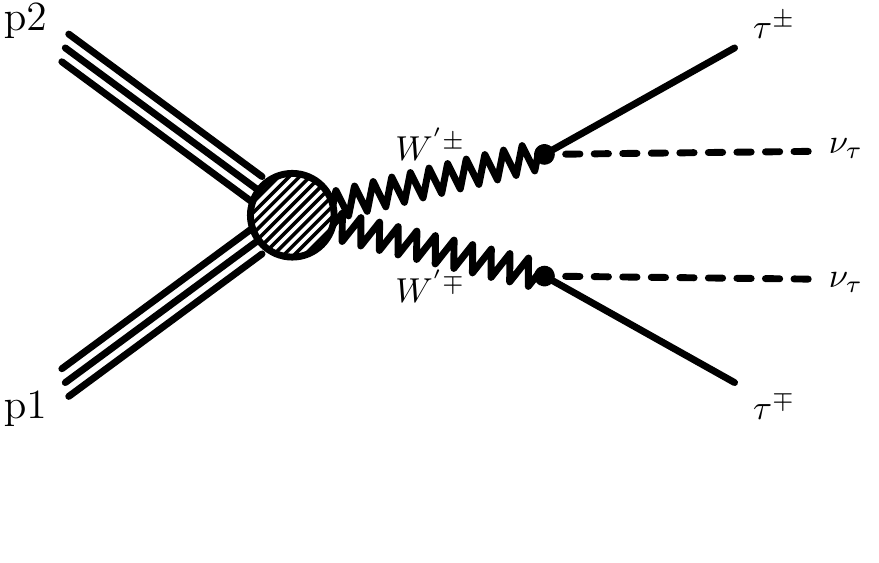}
	\caption{The diagram showing the production process of two \wprime bosons in collision of two protons and their decay into $\tau$ and $\nu_{\tau}$. The $\tau$ lepton decays either to lighter leptons, i.e. electron and muon, or to hadrons. For the analysis presented in this paper, at least one of the $\tau$ leptons decay hadronically.}
	\label{fig:wprimefeyndiagram}
\end{figure}

In this analysis, the efficiencies provided by the CMS experiment \cite{Khachatryan:2016trj} are used to find the yields of the favorite signal and compare it with the reported SM backgrounds to set a lower limit on the mass of \wprime boson. 
The CMS analysis uses LHC data from proton-proton (pp) collisions at a center-of-mass energy ($\sqrt{s}$) of 8 TeV to search for new physics in di-tau final states.  The data corresponds to an integrated luminosity of 18.1 and 19.6 $fb^{-1}$ in different channels. Three different final states are considered depending on the decay of two $\tau$ leptons, fully hadronic (\tauTau), where both $\tau$ leptons decay hadronically and \lepTau  ~(e\Tau ~or $\mu\Tau$), where one $\tau$ lepton decays hadronically and the other decays leptonically. The schematic diagram of decay is shown in figure \ref{fig:wprimefeyndiagram}.

In different experiments, many searches are done to see the signatures of \wprime boson, but up to now none of them have obtained a positive signal.
The most stringent limit is set by the ATLAS experiment in an analysis which looks at the tail of the transverse mass distribution of the lepton plus missing transverse momentum system \cite{Aaboud:2017efa}. The lepton is assumed to be produced in the decay of a \wprime boson associated with missing transverse momentum coming from a neutrino. It has excluded the \wprime boson with masses smaller than 5.1 TeV at a 95\% confidence level (CL), assuming a pure left-handed \wprime boson with \gL ~equal to the coupling of the SM $W$ boson (\gSM = 0.64).  The results of different direct searches from the colliders are also used to constrain the \wprime boson. For example, the results of the search for single top quark production are used to constrain the \wprime boson in Ref. \cite{YaserAyazi:2017xyj}.

In next section, the reference experimental analysis is reviewed and the used variables are defined. In Section \ref{sec:simulation}, the framework of our analysis is described. The results of the analysis are reported in Section \ref{sec:results}. Section \ref{sec:conclusion} concludes the paper.

\section{Review of the experimental analysis}
This study reinterprets the results of the ``Search for electroweak production of charginos in final states with two tau leptons in pp collisions at sqrt(s) = 8 TeV''\cite{Khachatryan:2016trj} for the production of two \wprime bosons decaying into two $\tau$ leptons. In this section the experimental analysis is reviewed.

The $\tau$ lepton decays  to a muon or an electron in $\sim 35\%$ of the cases and to hadrons (\Tau) in the rest of cases. So about 90\% of events with two $\tau$ leptons decay to \lepTau ~or \tauTau. In this analysis, only these two decay channels are considered and the fully leptonic decay modes are ignored. 

The missing transverse momentum (\MET) which is  defined as the vectorial sum of the transverse momenta (\pt) of neutrinos in the event is one of the event properties used in this analysis. In addition to the neutrinos  produced from the direct decay of \wprime boson, those produced in the $\tau$ lepton decay are also considered.

Having the momentum of the decay products of $\tau$ leptons and \MET, we can calculate all the needed variables used in the reference analysis. Stransverse mass ($M_{T2}$)~\cite{Lester:1999tx,Barr:2003rg}  is the main variable that is used to categorize the events. It is a function of momentum of two visible particles and \MET ~in the event. It is defined as:
\begin{equation}
M^2_{T2}(m_{N},\alpha,\beta,\MET) = \displaystyle\min_{p_T+q_T=\MET} [\max[M^2_{T}(\alpha,p),M^2_{T}(\beta,q)]]
\end{equation}
Where $\alpha$ and $\beta$ ($p$ and $q$) are the four momenta of the visible (invisible) decay products in two different legs and  $m_N$ is the mass of the invisible particle which is set to zero for this study. The transverse mass ($M_{T}$) is defined as :
\begin{equation}
M^2_{T}(\alpha,p) =  m^2_{\alpha}+m^2_N+2(E_T(p)E_T(\alpha)-\vec{p}_T.\vec{\alpha}_T)
\end{equation}
and transverse energy is given by; 
\begin{equation}
E_T(p)=\sqrt{p^2_T+m^2_N}
\end{equation}

In the reference analysis, the events are categorized in 4 signal regions (SR). For the e\Tau ~and $\mu\Tau$, it has been found that cutting the events with $\mttwo <$ 90 GeV is useful to discard the SM background events. But for \tauTau ~events, two separate SR's are defined as events with $\mttwo>$ 90 GeV (SR1) and events with 40 $<\mttwo<$ 90 GeV  and \SumMT $>$ 250 GeV (SR2), where \SumMT is the sum of the transverse mass of two \Tau ~objects.

In addition to the selection criteria that are applied for the lepton selection and event categorization, some other requirements to veto any extra lepton or b-tagged jets  are also applied. 

\section{Event generation}\label{sec:simulation}
To generate the signal events version 2.6.0 of   MadGraph5\_aMC@NLO~\cite{Alwall:2014hca} package is used which is the extension of MadGraph5~\cite{Alwall:2011uj} matrix-element generator. In the following, we refer to this event generator  as MadGraph.
The model used for signal generation is \wprime effective model (WEff-UFO)~\cite{Sullivan:2002jt}, which  is an extension of SM by adding the \wprime boson interactions with the SM fermions (Eq. \ref{eq:lagrangian}). The signal of our interest is $ pp\rightarrow \wprimep \wprimem$. 
To avoid the violation of unitarity at high energy, one needs to add interaction of \wprime boson with the SM gauge bosons to Eq. \ref{eq:lagrangian} and consider also the process $pp\rightarrow Z/\gamma^{\star} \rightarrow \wprimep \wprimem$. In this analysis, this part is ignored, because the goal is to study the leptonic interactions of \wprime boson, without adding the complexities introduced from the gauge interactions. It is checked and found that the ignored terms can increase the cross section of the favorite process by about 50\%, so the reported limits are conservative.

For each set of parameters, at least 20000 events are generated in pp collisions at $\sqrt{s}$ = 8 TeV. 
The momentum distribution of the partons in the proton is provided by the NN23LO1 \cite{Ball:2013hta} parton distribution function (PDF). The TAUOLA package~\cite{Davidson:2010rw} is used to simulate the $\tau$ lepton decays. It simulates the hadronic and leptonic decays of the $\tau$ lepton and provides full information about final state particles including neutrinos and mediator particles. It also considers spin information of the decay products in simulating the angular distribution of the decay products.

In the first study, we considered purely left-handed  \wprime (\gL = \gSM ~and \gR  = 0). 
This means interactions with quarks and leptons are both allowed. Different masses of  \wprime boson in the range of 100 through 400 GeV are used in this analysis.  
Decay width or life time of \wprime boson depends on decay modes, coupling strength of the decay process, and kinematic constraints. Decay widths corresponding to each mass of \wprime boson are estimated  by MadGraph. The results agree with the values reported in Ref.\cite{Sullivan:2002jt}, where the total width of \wprime and partial width of $\wprime \rightarrow t \bar{b},\bar{t}b $ are calculated in the leading order and the next to leading order precision.
The production cross section and total width for the decay of \wprime  to both quarks and leptons in different masses are listed in table \ref{tab:Xsec,L-h}. 
\begin{table}[htb]
	\centering
	\caption{Cross sections and decay widths when \gR = 0, \gL = \gSM = 0.64  for various \wprime masses for pp collisions at $\sqrt{s}$ = 8 TeV. \label{tab:Xsec,L-h} }
	\begin{tabular}{|c|c|c|}
		\hline 
		\wprime mass (GeV)  &  Decay width (GeV)  &  Cross section (fb)\\
		\hline 
		100 & 2.51 & 929 \\
		130 & 3.26 & 315 \\
		160 & 4.01 & 130 \\
		190 & 4.83 & 59.7 \\
		220 & 5.88 & 28.0 \\
		250 & 6.98 & 14.1 \\
		280 & 8.09 & 7.68 \\
		310 & 9.20 & 4.43 \\
		340 & 10.3 & 2.67 \\
		370 & 11.4 & 1.67 \\ 
		400 & 12.5 & 1.08 \\
		\hline
	\end{tabular}
\end{table}
The relative uncertainties on the cross sections reported in this table and the following tables are typically, 2-8\% from the scale variation and 3-5\% from PDF variation for different masses.

As a cross check the branching ratio (BR) which is defined as the partial decay width to a special channel divided by the total decay width is compared for both signs of the \wprime boson. 
It can be observed in table \ref{tab:W'BR} that the values are consistent for \wprimep ~and \wprimem ~bosons. 
\begin{table}[htb]
	\centering
	\caption{Branching ratios of \wprime  when \gR = 0, \gL = \gSM = 0.64 for various signs and masses of \wprime boson. \label{tab:W'BR} }
	\begin{tabular}{|c|c|c|c|c|}
		\hline 
		& \multicolumn{4}{c|}{Branching ratio}\\\cline{2-5}
		\wprime Mass (GeV) &   \wprimep $\rightarrow \bar{\tau},\nu_\tau $&   \wprimep$\rightarrow  t \bar{b}$ &   \wprimem$\rightarrow \tau,\bar{\nu}_\tau $ &  \wprimem$\rightarrow  \bar{t}b $ \\
		\hline 
		100  & 0.111  & 0.00   & 0.111  & 0.00\\
		190  & 0.110  & 0.0138 & 0.110  & 0.0138\\
		310  & 0.0939 & 0.155  & 0.0939 & 0.155\\
		400  & 0.0895 & 0.194  & 0.0895 & 0.194\\
		\hline
	\end{tabular}
\end{table}

\begin{figure}[htb]
	\centering
	\includegraphics*[width=.45\textwidth]{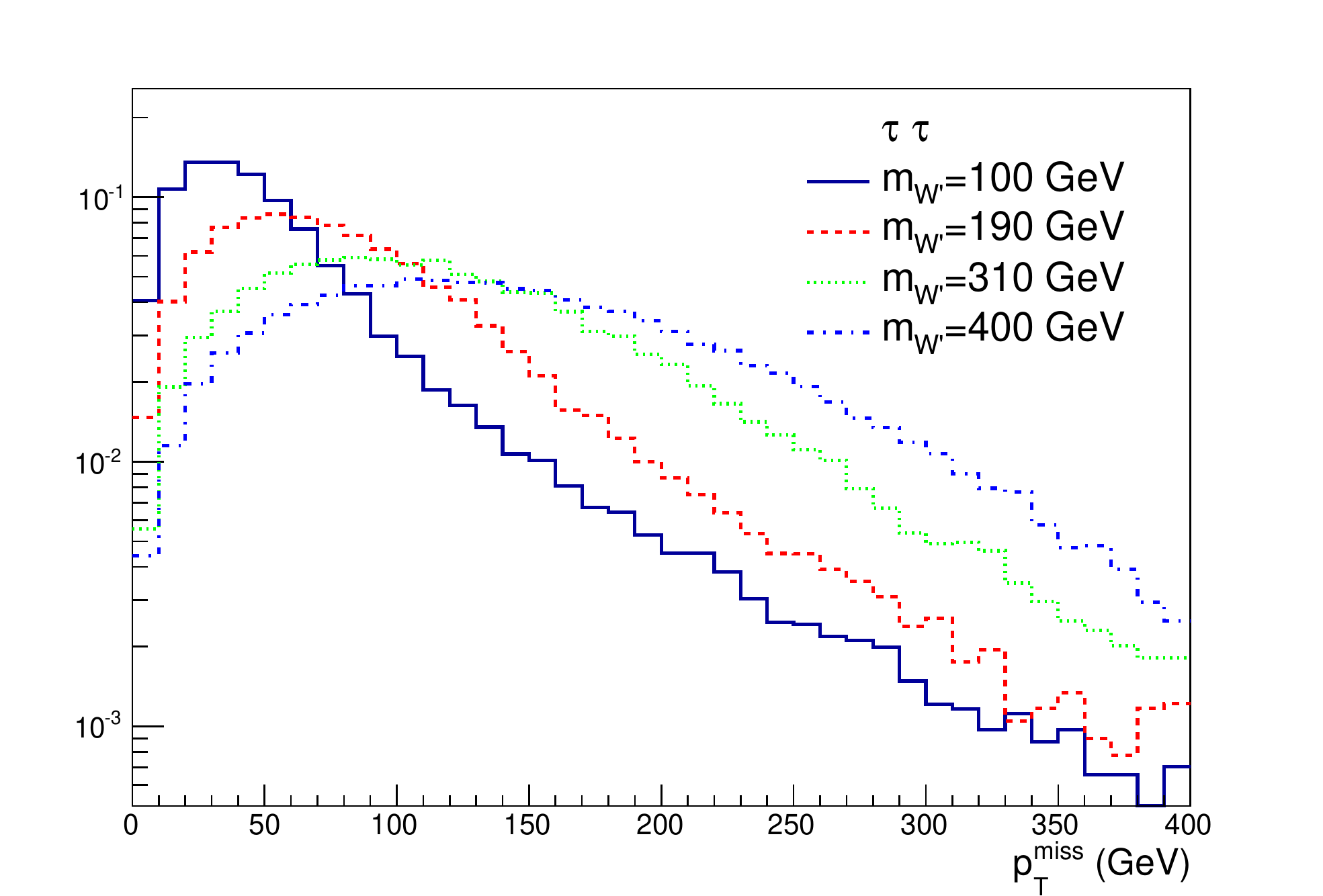}
	\hspace{3mm}
	\includegraphics*[width=.45\textwidth]{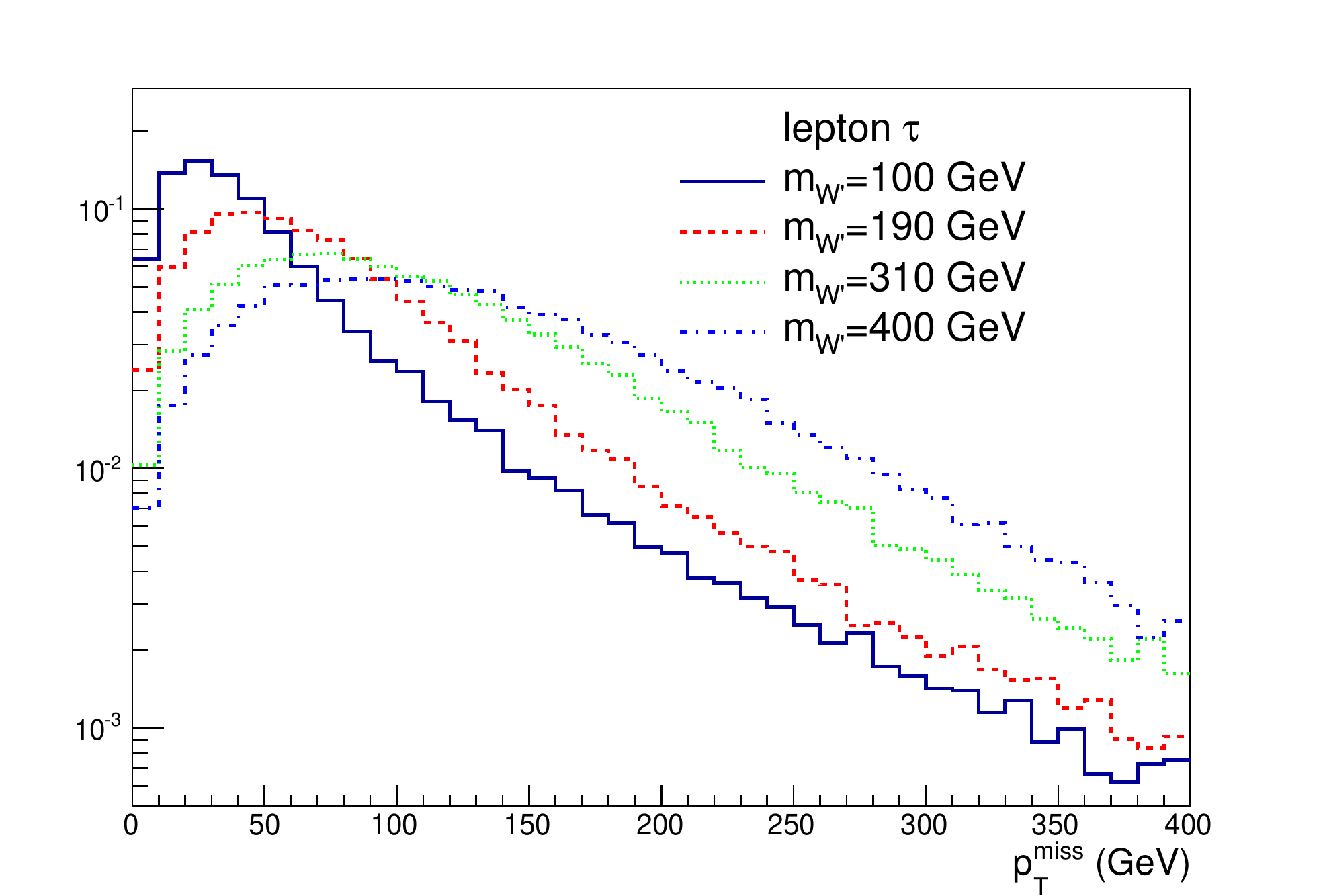}
	\caption{Missing transverse momentum (\MET) for different masses of \wprime boson. The events of \tauTau(\lepTau) channel are shown in left (right).}
	\label{fig:met}
\end{figure}

\begin{figure}[htb]
	\centering
	\includegraphics*[width=.45\textwidth]{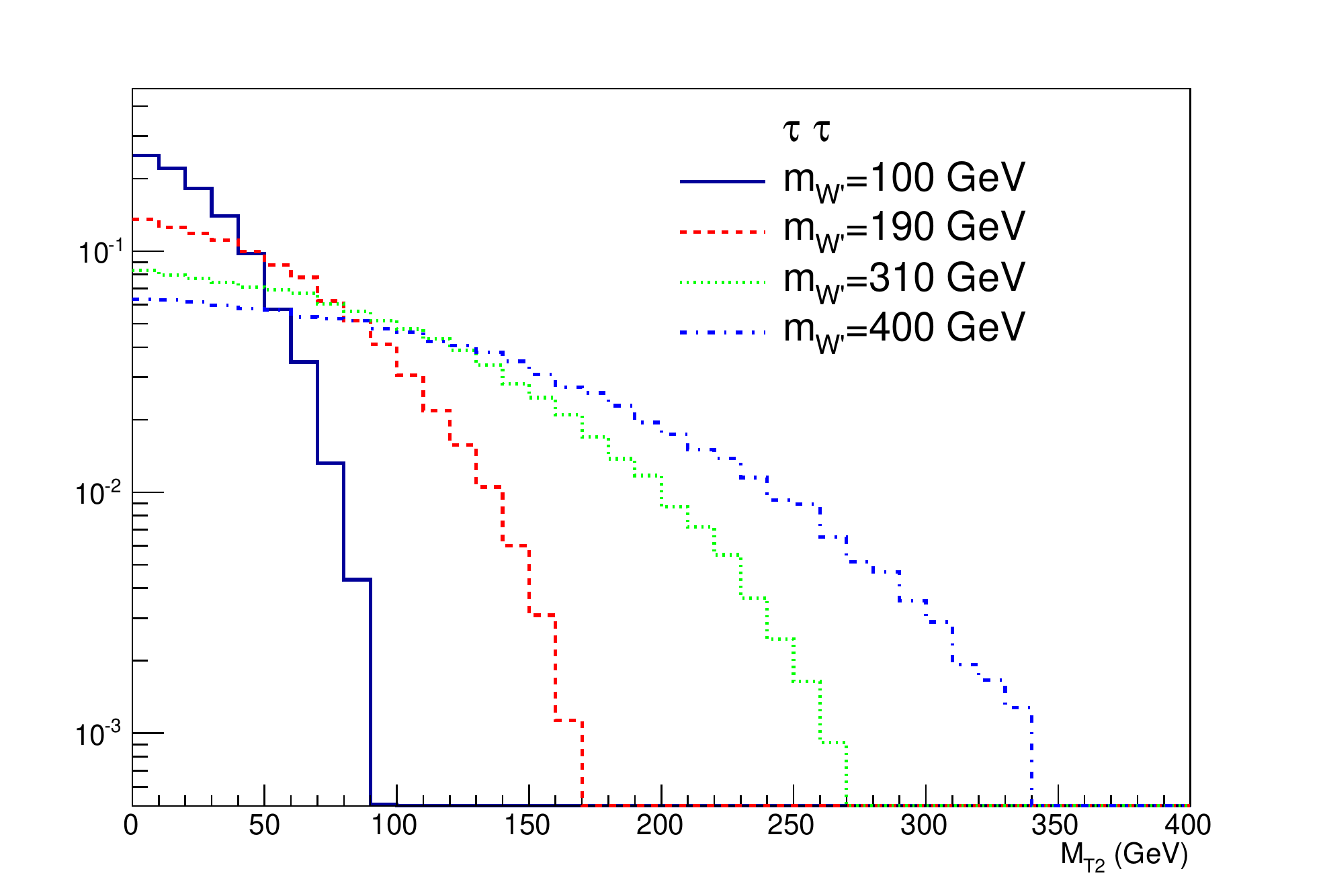}
	\hspace{3mm}
	\includegraphics*[width=.45\textwidth]{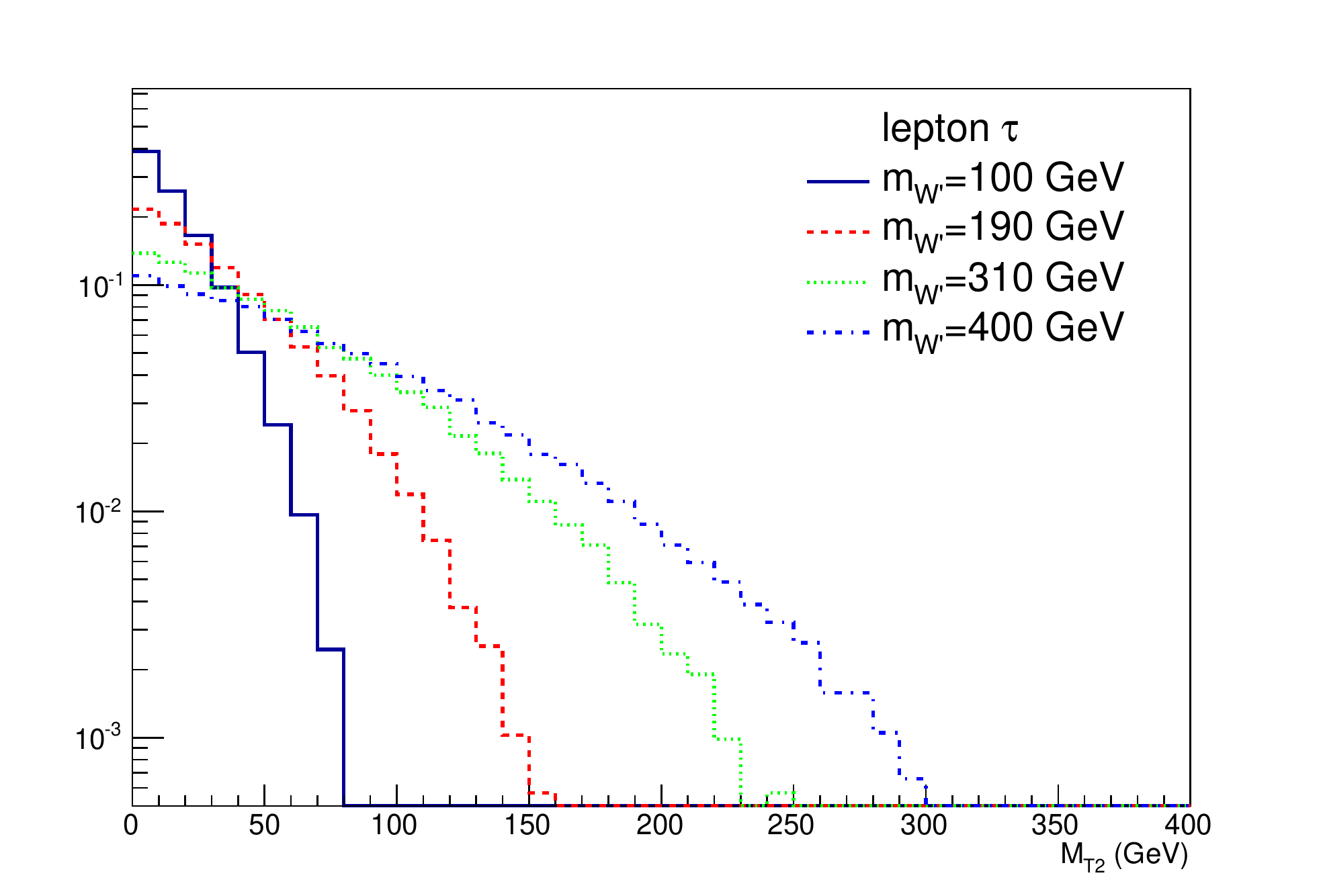}
	\caption{\mttwo ~for different masses of \wprime boson. The events of \tauTau (\lepTau) channel are shown in left (right).}
	\label{fig:mt2}
\end{figure}

\begin{figure}[htb]
	\centering
	\includegraphics*[width=.45\textwidth]{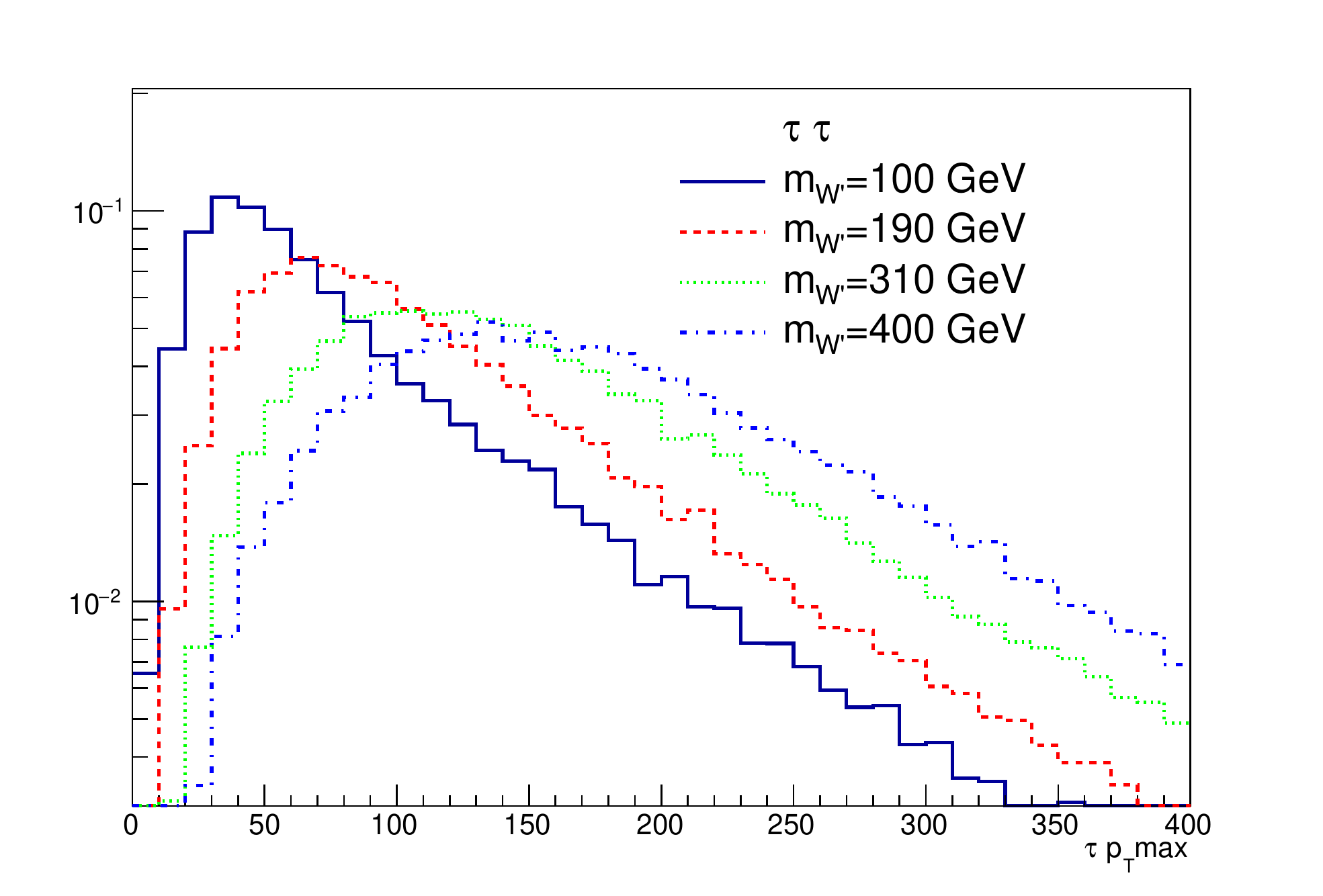}
	\hspace{3mm}
	\includegraphics*[width=.45\textwidth]{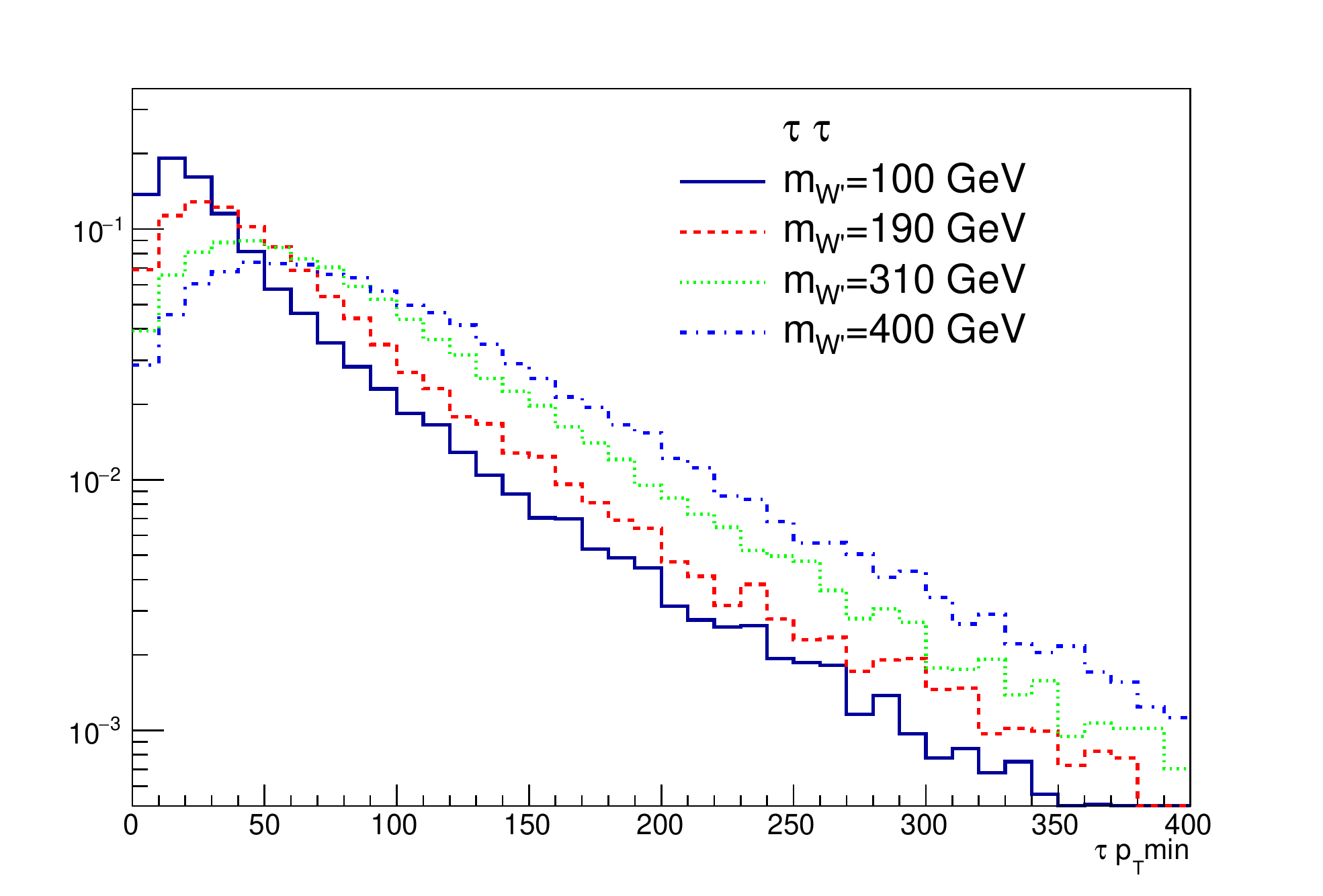}
	\caption{The maximum  and minimum of $\pt^{\Tau}$ in \tauTau ~channel for different masses of \wprime boson.}
	\label{fig:pt-hh}
\end{figure}

\begin{figure}[htb]
	\centering
	\includegraphics*[width=.45\textwidth]{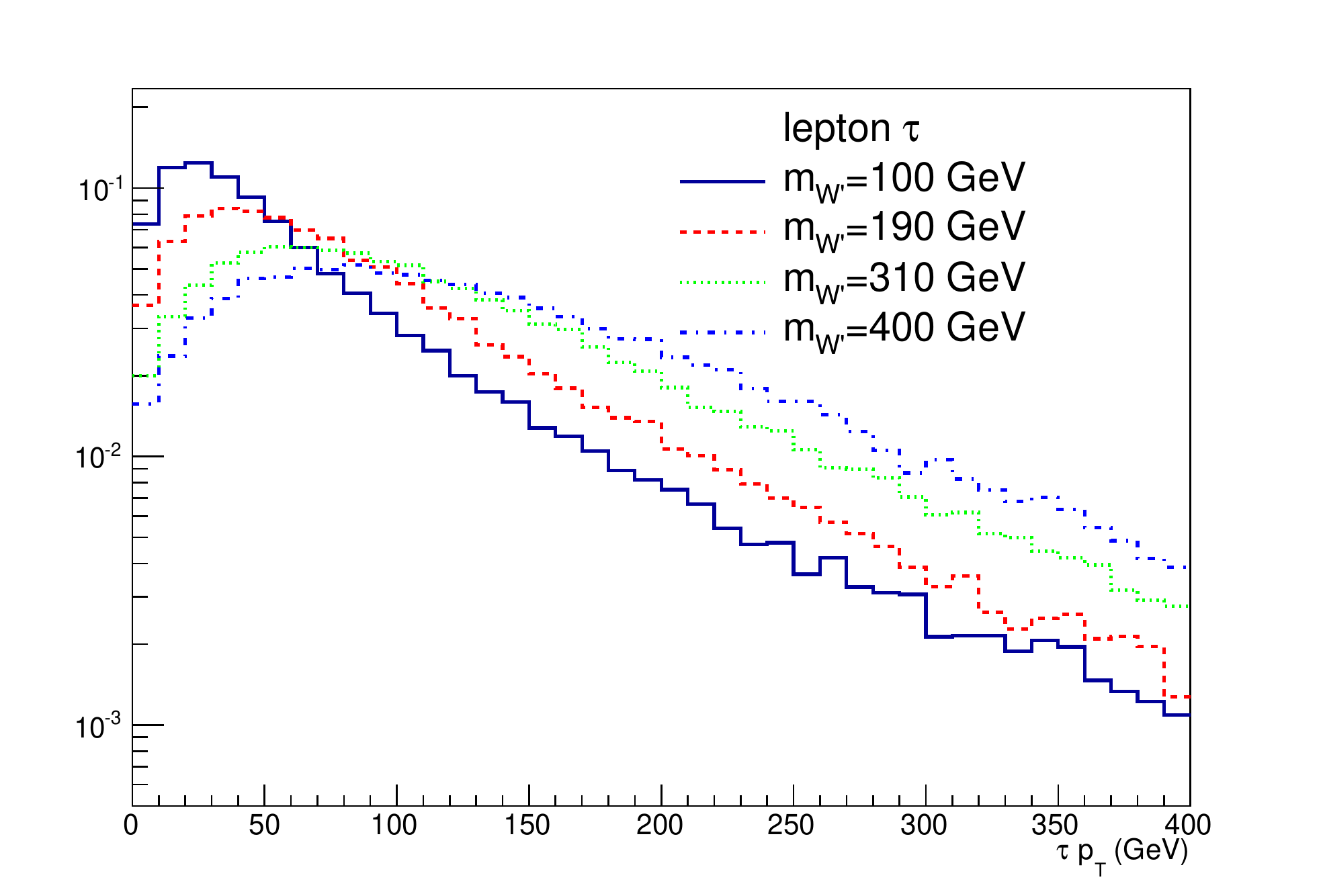}
	\hspace{3mm}
	\includegraphics*[width=.45\textwidth]{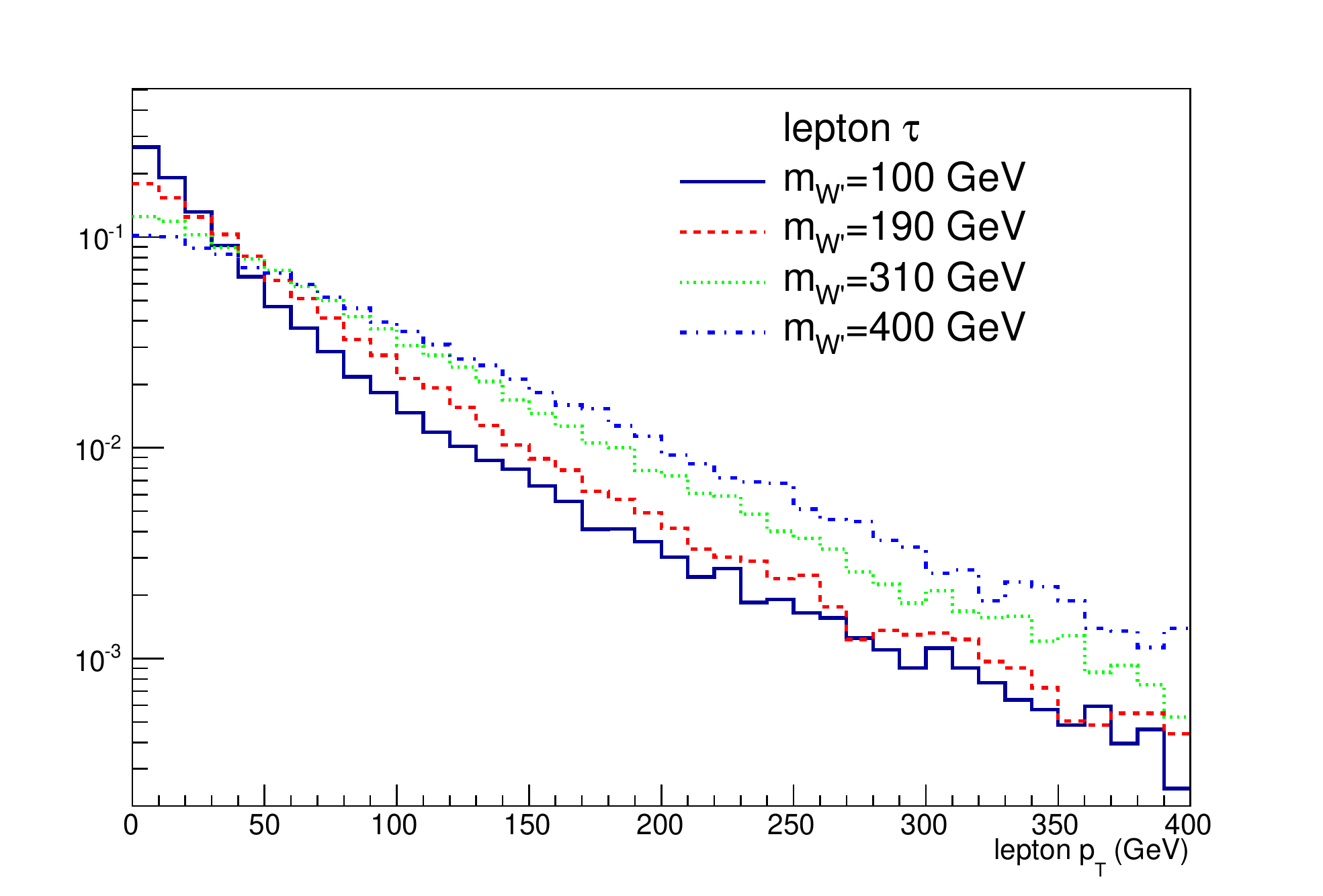}
	\caption{Left (right) plot shows $\pt^{\Tau}$ ($\pt^{\ell}$) in \lepTau ~channel for different masses of \wprime boson.}
	\label{fig:pt-lh}
\end{figure}

As another cross check, the kinematic  and search  variables of the generated events are produced. 
For this purpose, the visible momentum of the hadronic decaying $\tau$ is defined as the original $\tau$ momentum before decay subtracted by the momentum of the neutrinos in the decay chain of the $\tau$ lepton. The negative of the vectorial sum of the visible $\vec{p}_T$ ~of the two $\tau$ leptons defines \MET. Having these information, one can construct all the needed variables like the transverse mass of the leptons or \mttwo. 
As it is discussed earlier, the final state of our signal includes pure hadronic channel (\tauTau) and also a mixture of hadronic-leptonic channel (\lepTau ).  In figures \ref{fig:met} and \ref{fig:mt2}, the distributions of \MET ~and \mttwo ~for both channels in different \wprime masses are shown.
The transverse momentum of the leading and next-to-leading \Tau ~leptons in \tauTau ~channel are shown in figure \ref{fig:pt-hh}. The figure \ref{fig:pt-lh} shows the \pt ~of the lepton and \Tau ~in \lepTau ~channel. All of the distributions show the correct treatments and harder objects are produced when the mass of the \wprime boson is increased.

The couplings of the \wprime boson are not fixed by the model, so to investigate the effect of the couplings, we calculate the production cross section and decay width when the couplings are multiplied by 1.5 or 0.5. As can be seen in table \ref{tab:XsecgLVar}, 
\begin{table}[htb]
	\centering
	\caption{Cross sections and decay widths when \gL ~is decreased or increased by 50\% for various \wprime masses. \label{tab:XsecgLVar} }
	\begin{tabular}{|c|c|c|c|c|}
		\hline 
		& \multicolumn{2}{c|}{\gR = 0, \gL = $\frac{1}{2}$\gSM = 0.32}
		& \multicolumn{2}{c|}{\gR = 0, \gL =$\frac{3}{2}$\gSM = 0.96}\\\cline{2-5}
		\wprime Mass (GeV)  &  Decay width (GeV) &  Cross section(fb)&  Decay width (GeV) &  Cross section(fb)\\
		\hline 
		100& 0.627 & 58.1  & 5.50 & 4410\\
		130& 0.815 & 19.7  & 7.15 & 1500\\
		160& 1.00  & 8.16  & 8.80 & 618\\
		190& 1.21  & 3.72  & 10.6 & 284\\
		220& 1.47  & 1.75  & 12.9 & 132\\
		250& 1.75  & 0.885 & 15.3 & 67.1\\
		280& 2.02  & 0.482 & 17.7 & 36.3\\
		310& 2.30  & 0.278 & 20.2 & 21.0\\
		340& 2.58  & 0.168 & 22.6 & 12.0\\
		370& 2.85  & 0.105 & 25.0 & 7.48\\
		400& 3.11  & 0.0678& 27.3 & 5.11\\    
		\hline
	\end{tabular}
\end{table}  
the cross section is scaled by 5.06 and 1/16 when the coupling is increased and decreased by 50\%, respectively. It is noticeable that the values of decay widths are proportional to  factors of 2.25 and  0.25 for the increased  and decreased left-handed coupling values. The behaviors of the cross sections, $(\gL)^4$, and decay widths, $(\gL)^2$, are consistent with our expectations.

In an alternative approach, the left-handed and right-handed couplings are changed in a way that their squared sum is constant and equal to $\gSM^2$.
\begin{equation}
\gSM^2 = (\gL)^2 +  (\gR)^2 
\end{equation}
It is easier, to define a mixing angle and rewrite the couplings as:
\begin{eqnarray}
\gL  = \gSM \cos\theta \\
\gR  = \gSM \sin\theta
\end{eqnarray}
By varying $\theta$ from 0 to $ 90^\circ $, the \wprime goes from a purely left-handed to a purely right-handed vector boson. The latter \wprime boson does not have any interaction with the leptons. 
The variation of the cross sections and decay widths due to various mixing angles, when \wprime mass is 310 GeV, is shown in table \ref{tab:mixingAngle}.
\begin{table}[htb]
	\centering
	\caption{Cross sections and decay widths of different mixing angles for a 310 GeV \wprime boson. \label{tab:mixingAngle} }
	\begin{tabular}{|c|c|c|c|c|}
	\hline 
	Mixing angle $\theta$  & \gR,  \gL & Decay width (GeV)  &  Cross section (fb) & BR(\wprime $\rightarrow \tau \nu_\tau$) \\
	\hline 
	0$^\circ$  & 0.0, 0.64  & 9.20  & 4.43 & 0.0939 \\
	30$^\circ$ & 0.32, 0.56 & 8.51  & 1.78 & 0.0757\\
	45$^\circ$ & 0.46, 0.46 & 8.00  & 0.752 & 0.0543\\
	60$^\circ$ & 0.56, 0.32 & 7.21  & 0.263 & 0.0292\\
	\hline
\end{tabular}
\end{table}

In the next section the generated events in this section are used to set a lower limit on the \wprime mass.

\section{Results}\label{sec:results} 
Using the \wprime samples generated by MadGraph as explained in Section \ref{sec:simulation} and decaying $\tau$ leptons using the TAUOLA package, we are ready to measure the efficiency of the selection for different channels for different \wprime masses. 

For each event, the probability of passing the selection cuts for a given signal region can be obtained using the cut efficiency tables of the experimental paper \cite{Khachatryan:2016trj}. In that paper, the efficiency of applying each cut on the reconstructed properties of the event is reported as a function of the generator level value of that property. It makes it very easy and accurate to take into account the detector effects that are always difficult to model. Following that paper, all the cuts are considered independent.  The efficiencies of different cuts are multiplied to obtain the full selection efficiency for different channels and signal regions.

\begin{figure}[!htb]
	\centering
	\includegraphics*[width=.45\textwidth]{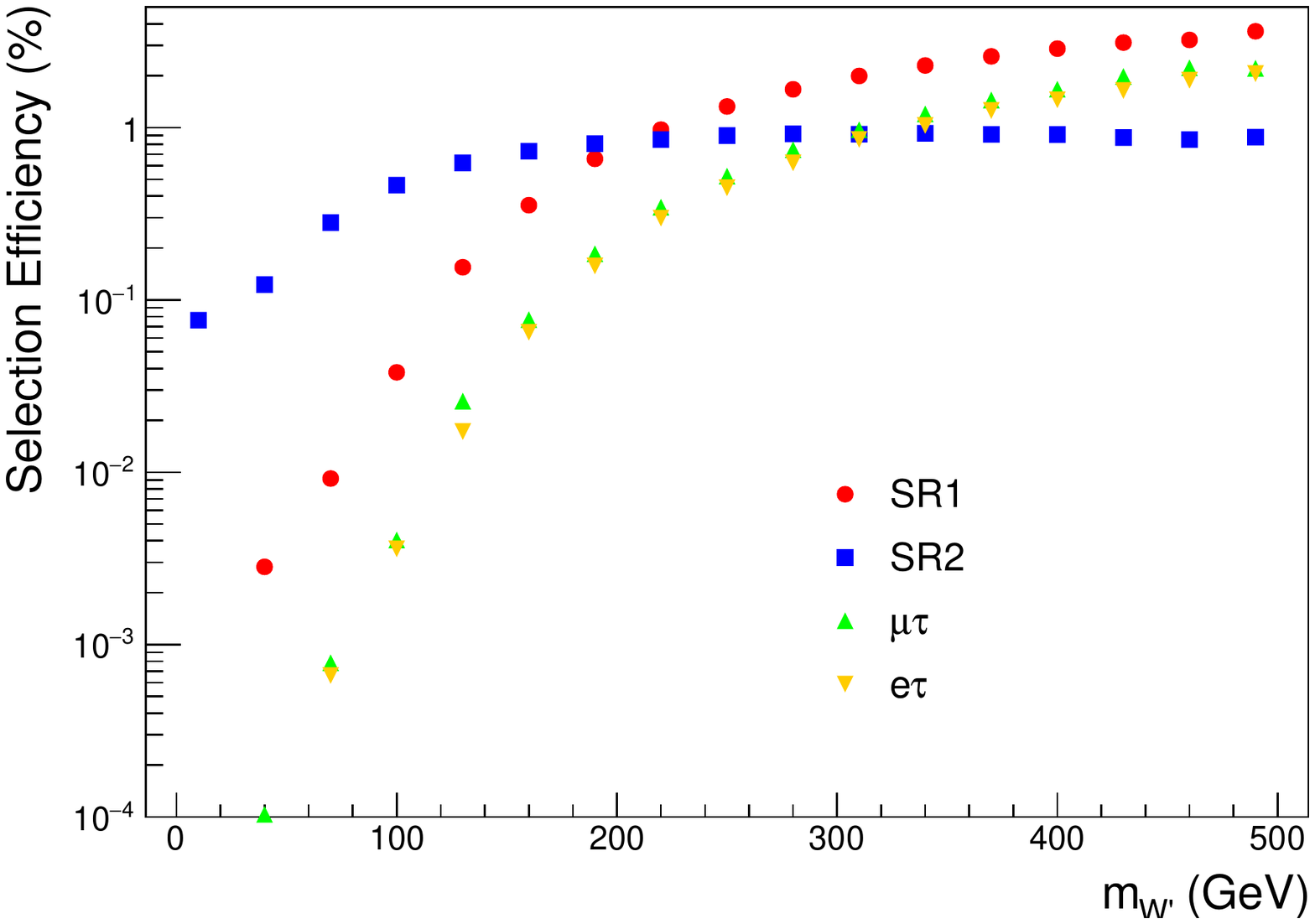}
	\caption{Efficiecny of signal selection in different signal regions as a function of the \wprime mass.}
	\label{fig:EfficiencyGraphs}
\end{figure}

This was done for different \wprime masses and for different coupling strengths. The resulting efficiencies for the SM-like scenario can be seen in figure~\ref{fig:EfficiencyGraphs}. 
The efficiencies for the case where \gL ~is increased or decreased by 50\% or where there is also a non-zero \gR ~are produced and compared with the results in figure~\ref{fig:EfficiencyGraphs}. As it is expected, the efficiencies depend only on the kinematic of the generated events which vary with the mass of the \wprime boson and do not depend on the coupling constants.

Having the full selection efficiency in one channel ($\varepsilon^{ch}_{fs}$), together with the production cross section ($\sigma$) and the decay branching ratio (BR), one can estimate the total number of expected signal events in a given integrated luminosity ($\mathcal{L}$) using the formula:
\begin{equation}
N^{ch}_{exp.}= \mathcal{L} \times \sigma(pp \to \wprime\wprime) \times BR^{2}(\wprime \to \tau \nu) \times \varepsilon^{ch}_{fs}
\end{equation}
According to the experimental paper, the integrated luminosity for the \tauTau ~signal regions is 18.1~fb$^{-1}$ and for the \lepTau ~channels is 19.6~fb$^{-1}$. Following the same reference, a systematic uncertainty of 20\% for signal in \lepTau ~channel and 25\% in \tauTau ~channel is assumed. Data yields and background predictions with their uncertainties in the four signal regions of search obtained from Ref.\cite{Khachatryan:2016trj} are shown in table \ref{tab:yields}. 
\begin{table}[htb]
	\centering
	\caption{Data yields and background predictions with their uncertainties for \lepTau ~and \tauTau ~channels. The shown uncertainty is the quadratic sum of the statistical and systematic uncertainties provided by the CMS experiment.\label{tab:yields} }
	\begin{tabular}{|c|c|c|c|c|}
		\hline 
		&    e\Tau       &  $\mu\Tau$     &  \tauTau ~SR1  & \tauTau ~SR2 \\
		\hline 
		Background &3.52 $\pm$ 3.39 &8.59 $\pm$ 4.83 &1.58 $\pm$0.65 &7.07 $\pm2.25$ \\     
		Observed data& 3            &      5         &    1          &    2    \\  
		\hline
	\end{tabular}
\end{table}

\begin{figure}[!htb]
	\centering
	\includegraphics*[width=.45\textwidth]{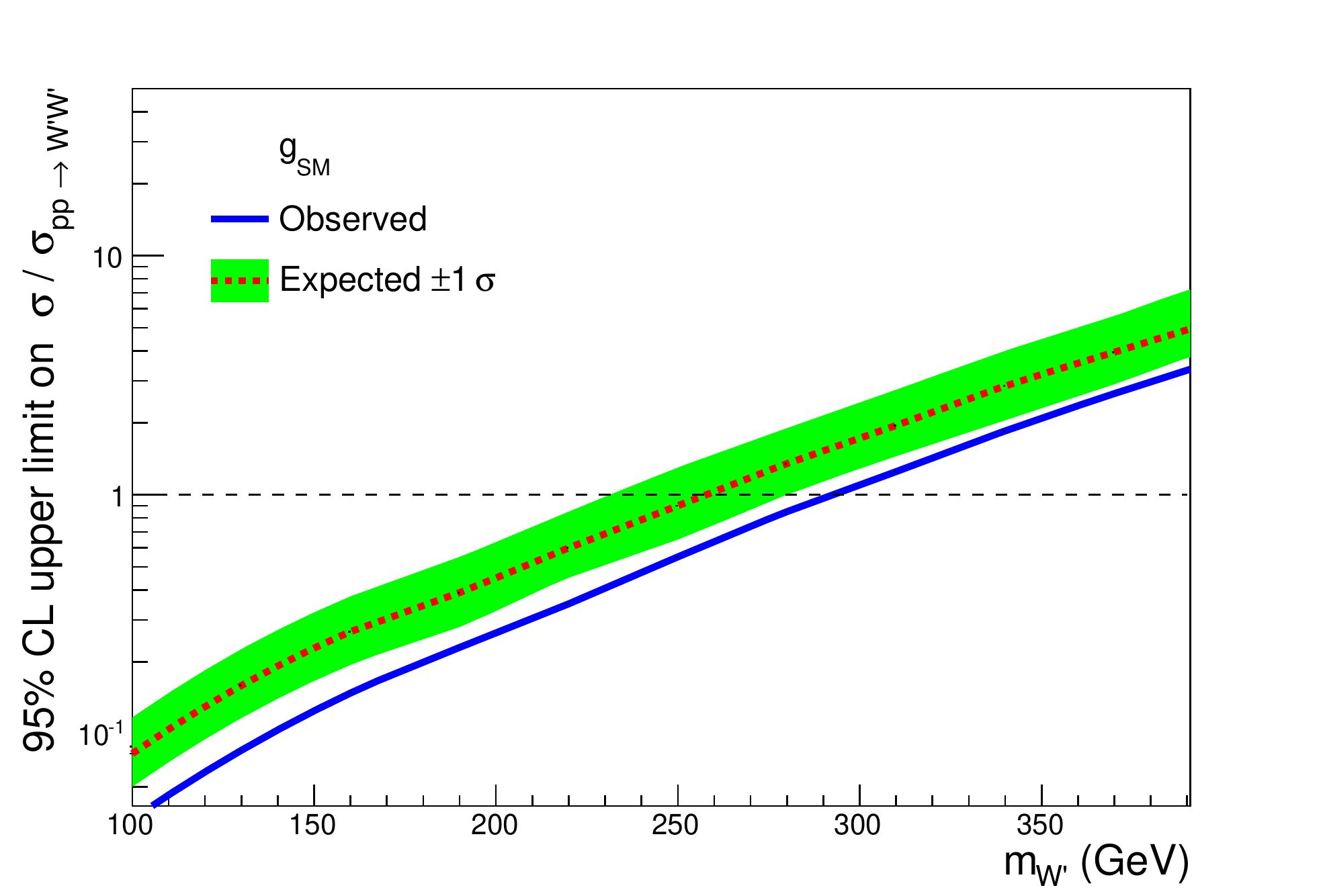}
	\vspace{3mm}	
	\includegraphics*[width=.45\textwidth]{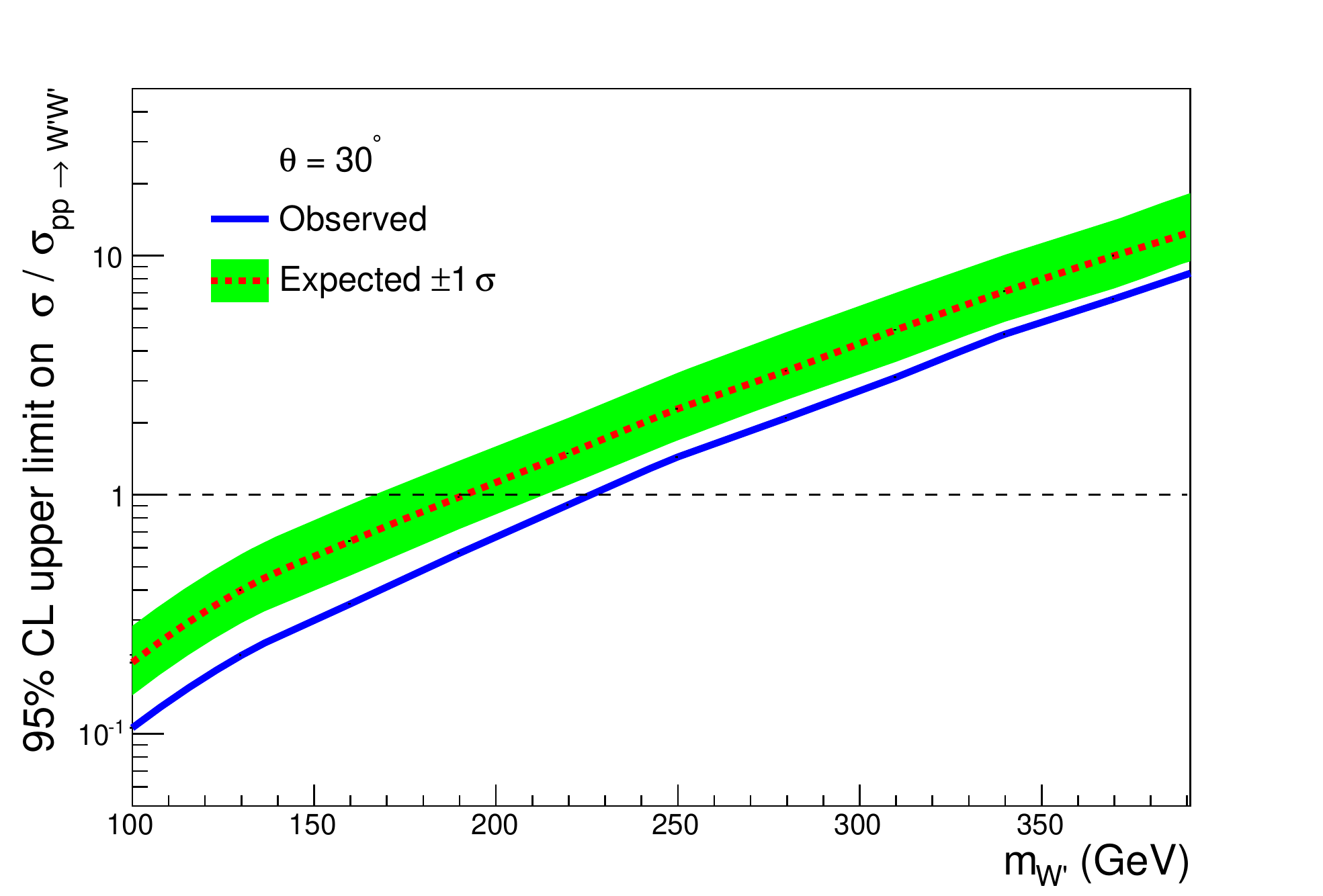}
	\hspace{3mm}
	\includegraphics*[width=.45\textwidth]{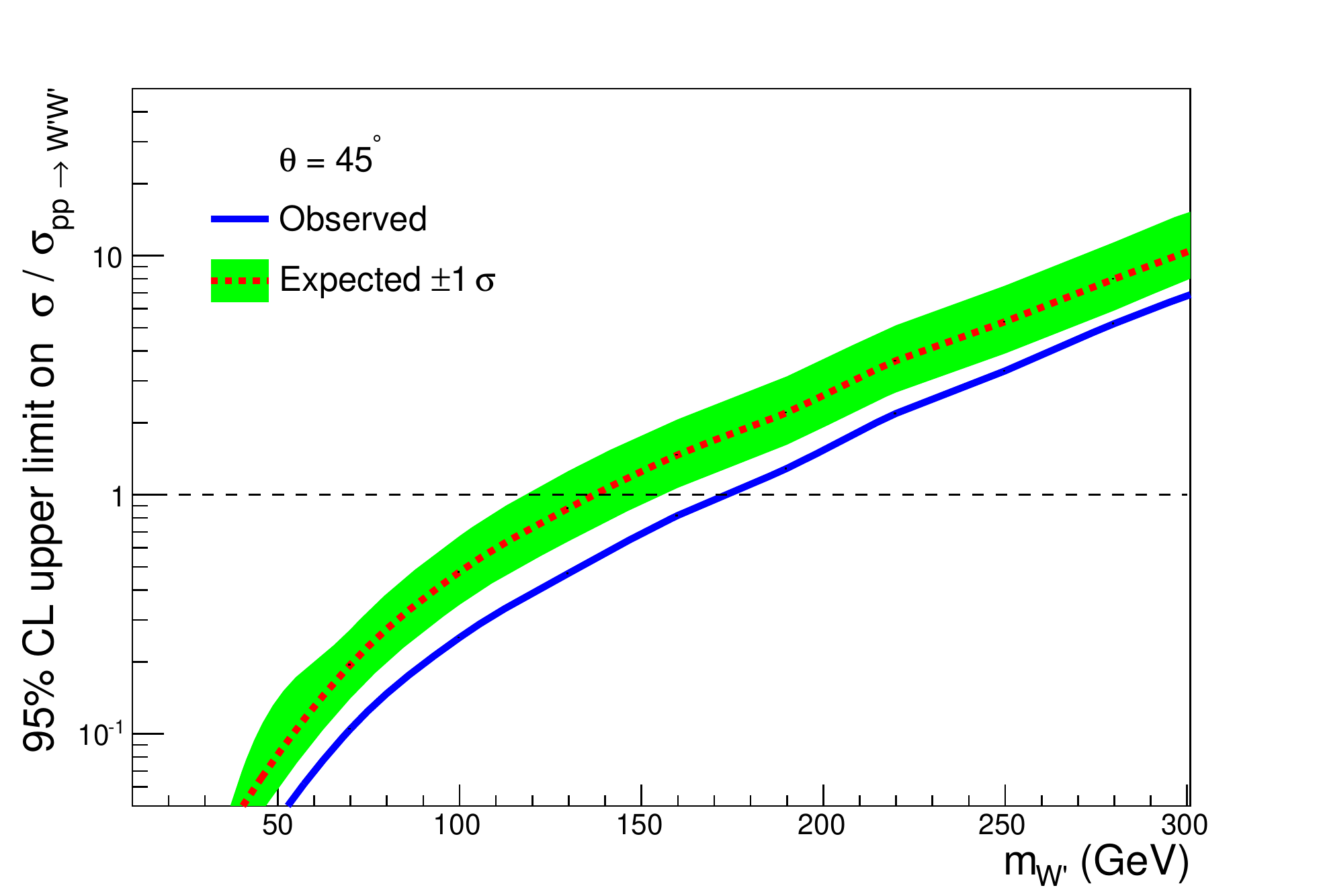}
	\vspace{3mm}	
	\includegraphics*[width=.45\textwidth]{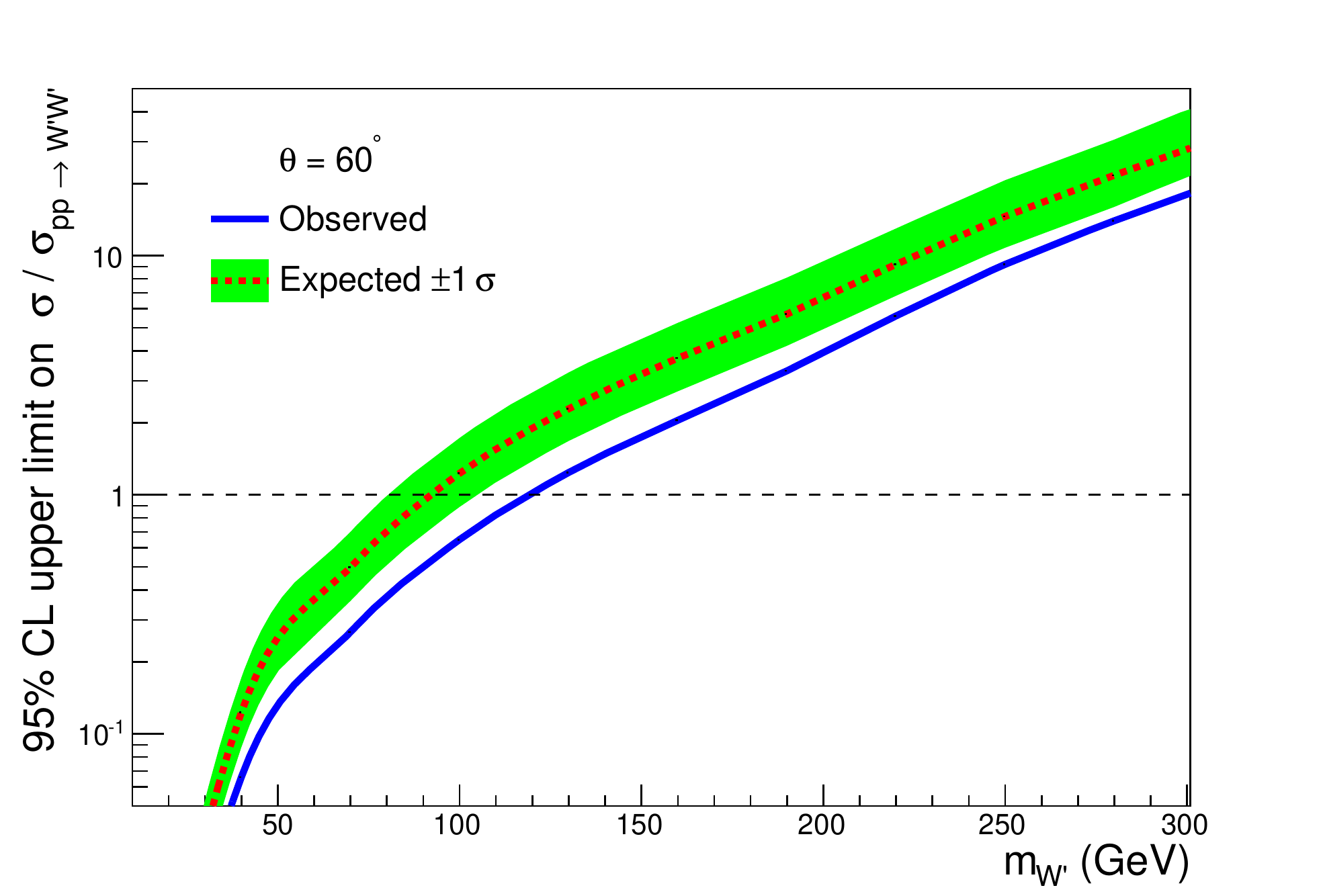}
	\hspace{3mm}
	\includegraphics*[width=.45\textwidth]{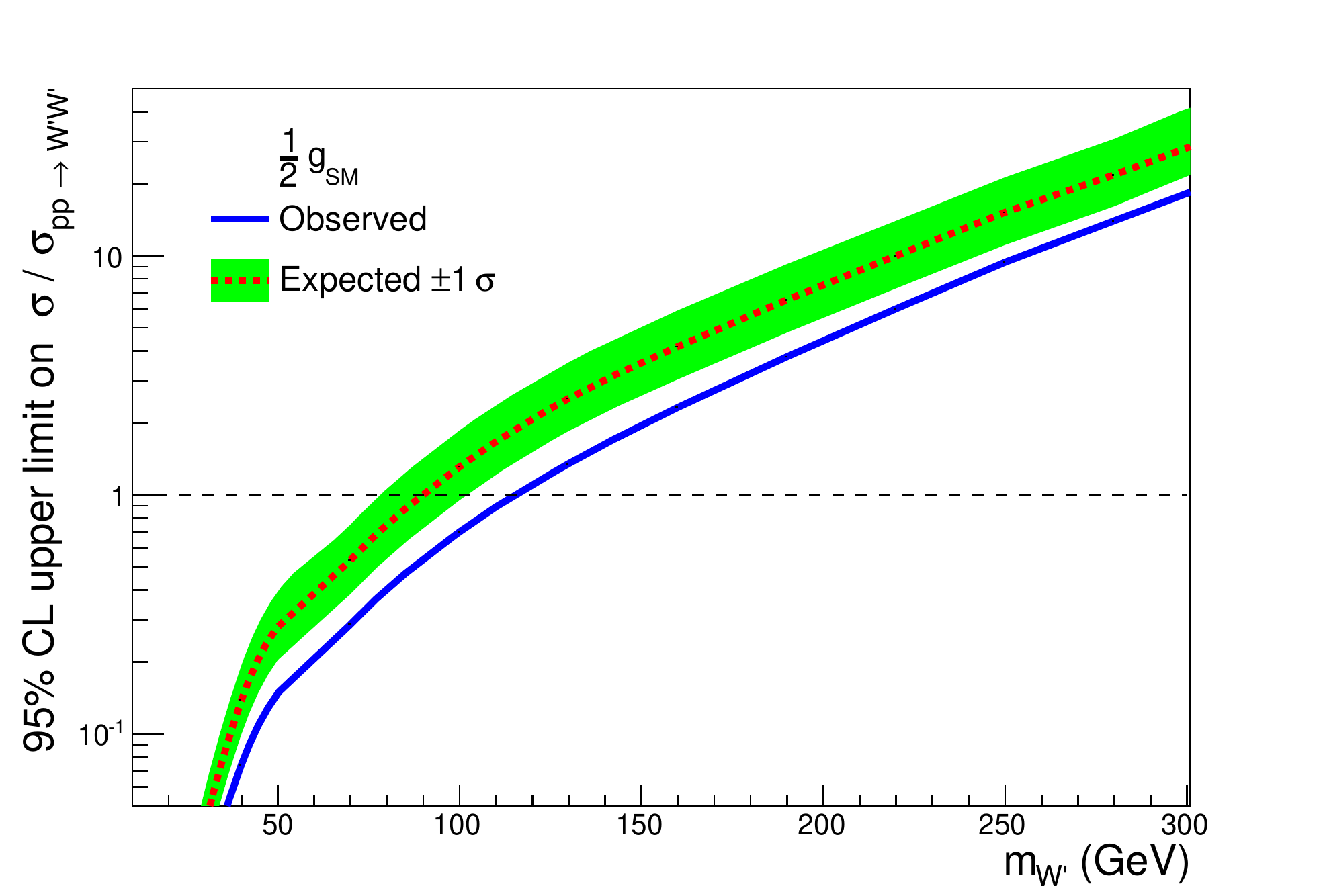}
	\vspace{3mm}
	\includegraphics*[width=.45\textwidth]{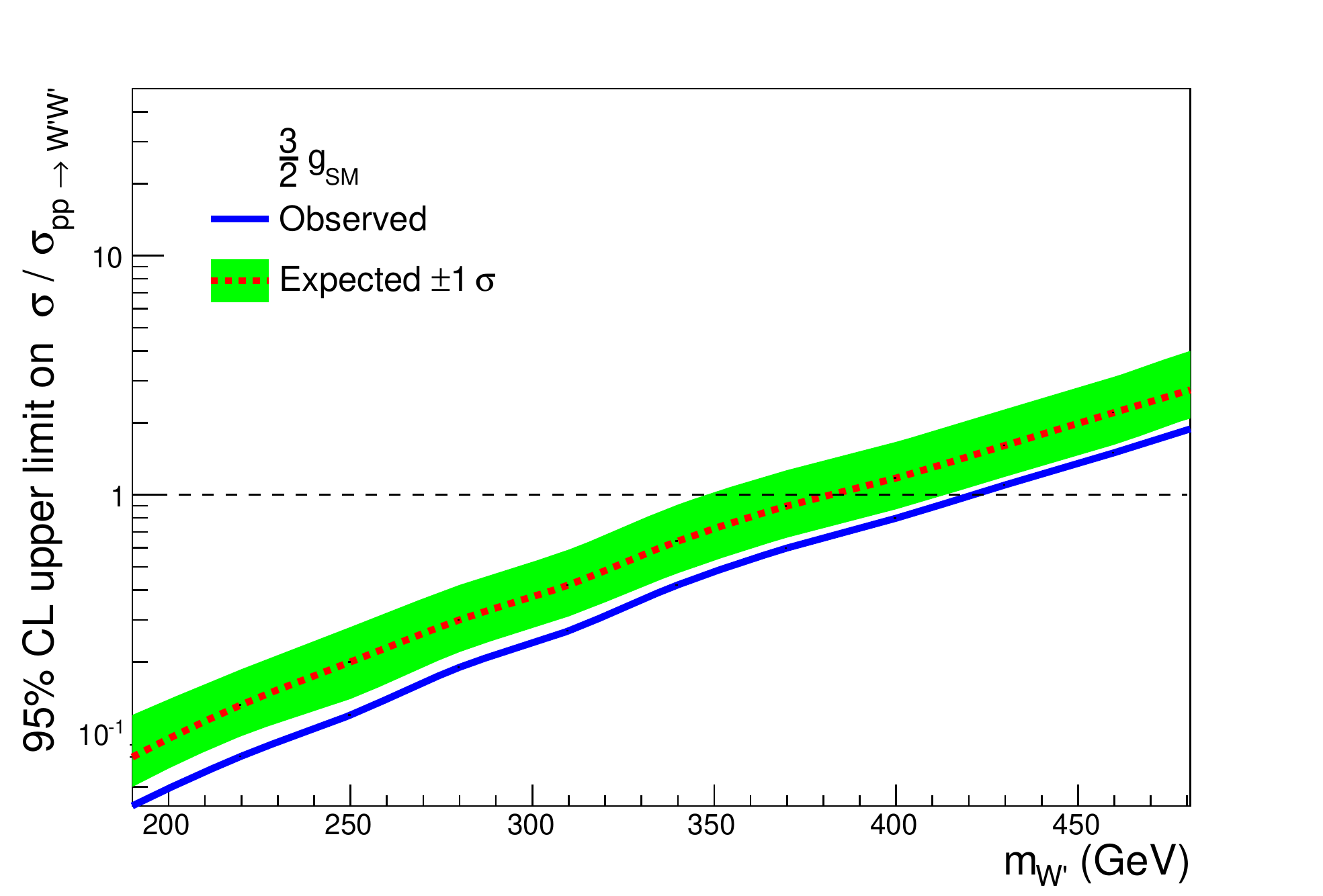}
	\caption{Upper limit of $\sigma/\sigma_{pp \rightarrow \wprime \wprime}$ production for different scenarios}
	\label{fig:brazilianFlags}
\end{figure}

The 95\% confidence level upper limit on the signal strength can be found by combining all the four channels. A Likelihood ratio semi-bayesian method implemented in ROOT \cite{Brun:1997pa} is used. Signal strength is defined as the $\sigma /\sigma_{pp \to \wprime\wprime}$ratio. Results for the SM-like \wprime are shown in figure  \ref{fig:brazilianFlags} (top-left). 
It can be seen that \wprime masses up to  290 GeV are excluded.  

Repeating this procedure for the other scenarios of the coupling constants, it is observed that when \gL ~is decreased, the sensitivity is decreased, and vice versa, as it is expected. Figure \ref{fig:brazilianFlags} shows the observed limits, the expected exclusions and $\pm 1 ~\sigma$ uncertainties on the expected exclusions for different scenarios of the coupling constants. Table \ref{tab:ObservedLimits}
\begin{table}[htb]
	\centering
	\caption{The expected and observed lower limits on the \wprime mass in different scenarios of the coupling constants. \label{tab:ObservedLimits} }
	\begin{tabular}{|c|c|c|}
		\hline
		\gR, \gL           & Expected (GeV) & Observed (GeV) \\\hline
		0, 0.64            &   255    &    290  \\
		0.32, 0.56         &   190    &    225  \\
		0.46, 0.46         &   135    &    170  \\
		0.56, 0.32         &   90     &    120  \\
		0, 0.32            &   90     &    120  \\
		0, 0.96            &   380    &    420  \\\hline
	\end{tabular}
\end{table}
summarizes the expected and observed limits in different scenarios. Always, the observed limit is higher than the expected one, because in different signal regions the observed data is less than the expected background (Table. \ref{tab:yields}).

The results are lower than the results from the direct search, but the proposed method can be used to constrain any new model with a similar final state, without need to simulate the response of a real detector.

\section{Conclusions}\label{sec:conclusion} 
With increasing the center of mass energy of the colliders, pair production of the heavy charged bosons, \wprime bosons, can be accessible. The event selection efficiencies, provided by a CMS analysis in a similar final state enables us to approximate the detector effects without the complexities from the full detector response simulation. The efficiencies are used to find the yield of the favorite signal. A statistical tool is used to compare the yield of the signal with the observed events from data and set a lower limit on the mass of the \wprime boson. Different 
scenarios for the coupling of \wprime bosons to leptons is examined and the corresponding lower limits are reported. If the coupling constants are same as those of the SM $W$ boson, the masses up to 290 GeV are excluded at a 95\% confidence level. Depending on the scenario, the limit can be lower or even pushed up to 420 GeV.

\section{Acknowledgments}
The analysis has benefited significantly from ideas and comments by Hamed Bakhshiansohi. We have to thank him for all of his helps and supports. 
The authors would like to thank Seyed Yaser Ayazi, for the useful discussions and 
his comments on the manuscript. The authors are grateful to CMS collaboration for their fantastic results.

\end{document}